\documentclass[journal  =  jpccck, layout=onecolumn]{achemso}
\usepackage{color}
\usepackage{amssymb}
\usepackage[english]{babel}
\usepackage{amsmath}
\usepackage{appendix}
\usepackage{amsthm}
\usepackage{amsfonts}
\usepackage[version=3]{mhchem} %chem package, \ce{Ag25L14} gives cluster size, evt + subscript, superscript
\usepackage{graphicx}
\usepackage{dcolumn}
\usepackage{bm}
\usepackage{subfigure}

\def\be{\begin{equation}}
\def\ee{\end{equation}}

\title{Chain-Assisted Charge Transport in Semicrystalline Conjugated Polymers}
\author{B.O. Conchuir}
\affiliation{Cavendish Laboratory, JJ Thomson Avenue, CB30HE Cambridge,
U.K.}
\alsoaffiliation{Institute of Particle Technology (LFG), Friedrich-Alexander University Erlangen-N\"{u}rnberg (FAU), Cauerstrasse 4, 91058 Erlangen, Germany.}
\email{breanndan.oconchuir@fau.de}  

\author{C. Tarantini}
\affiliation{Cavendish Laboratory, JJ Thomson Avenue, CB30HE Cambridge,
U.K.}

\author{C.R. McNeill}
\affiliation{Department of Materials Engineering, Monash University, Clayton Victoria,
Australia.}

\author{S. H\"{u}ttner}
\affiliation{Physical Chemistry, Universit\"{a}t Bayreuth, 95440 Bayreuth,
Germany.}

\author{A. Zaccone}
\affiliation{Cavendish Laboratory, JJ Thomson Avenue, CB30HE Cambridge, U.K.}
\alsoaffiliation{Department of Chemical Engineering and Biotechnology, University of Cambridge, New Museums Site, Pembroke Street,
CB2 3RA Cambridge, U.K.}

\begin{document}

\begin{tocentry}
\includegraphics[width=4.5cm]{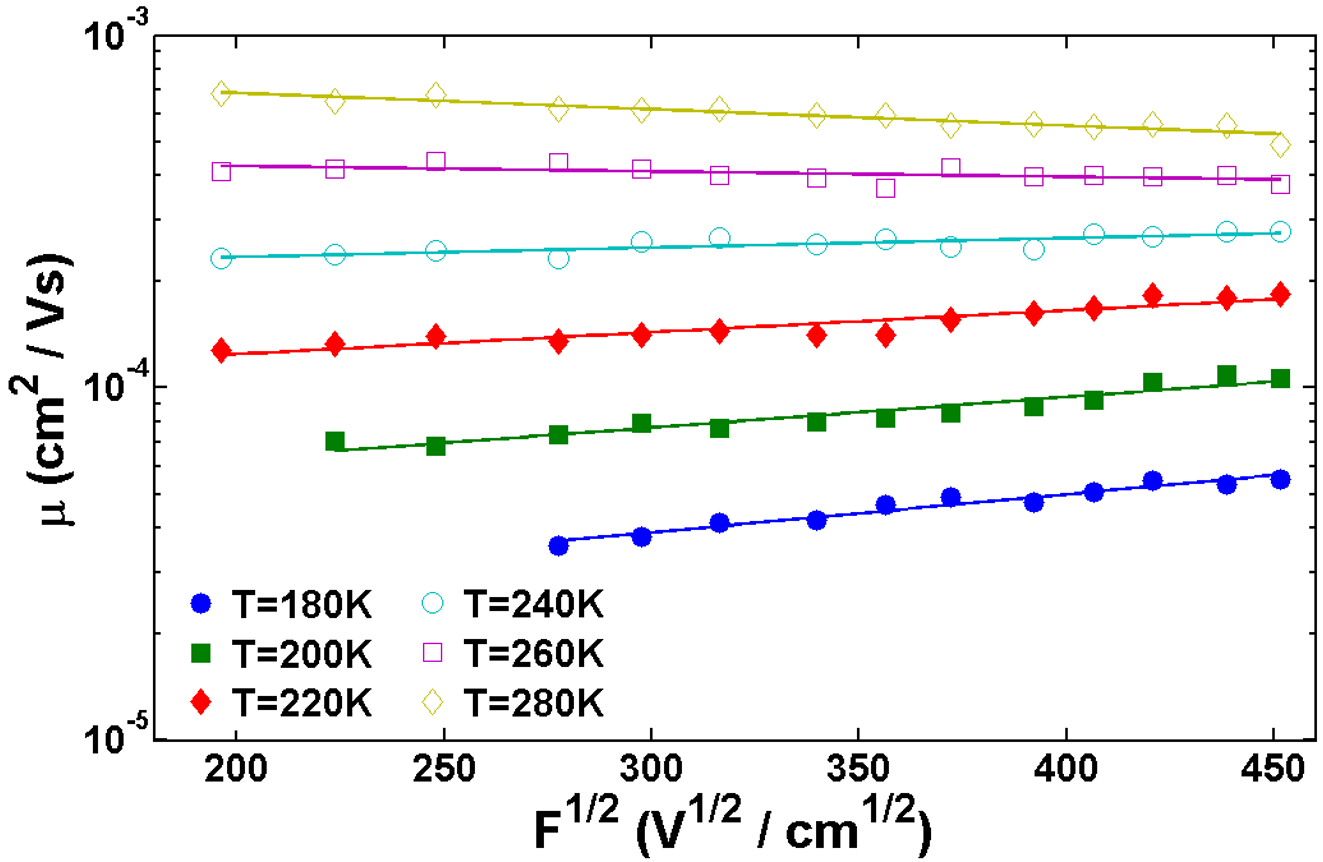}
\includegraphics[width=4.5cm]{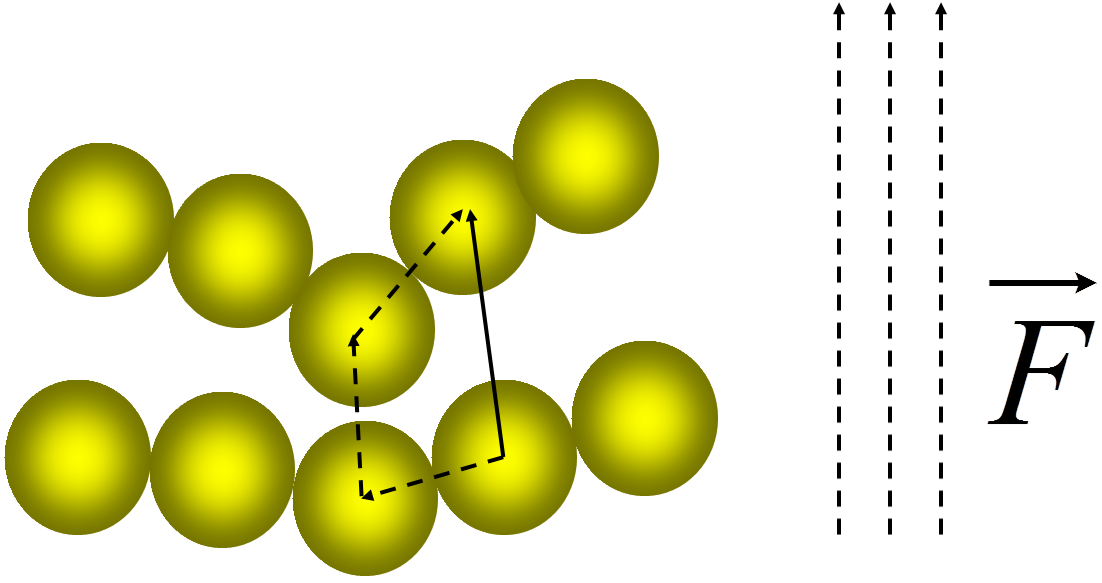}
\end{tocentry}

\begin{abstract}
{Charge-carrier transport in a paradigmatic semicrystalline polymer semiconductor (P3HT) is important for both fundamental understanding and applications. In samples with enhanced structural disorder due to ad-hoc point defects, the mobility displays rich behavior as a function of electric field ($F$) and temperature ($T$). At low $T$, the mobility increases with the applied field but upon further increasing T, the field-dependence becomes shallower. Eventually, at the highest T considered, the slope changes sign and the mobility then decreases with the field. This phenomenon can be interpreted with our model as a result of the competition between intra-chain conductive-like transport (which slows down on increasing $F$) and inter-chain activated transport (which is faster at higher $F$). The former is controlling at high T where inter-chain hops are strictly limited to nearest-neighbour monomers on adjacent chains. At low-T, instead, inter-chain hops to distant sites are allowed and control the positive correlation of the mobility with the field.} 
\end{abstract}

\section{Introduction}
The prospect of low-cost, flexible and efficient organic solar cells, organic field-effect transistors and organic light-emitting diodes has sparked an intensive effort to determine the charge transport properties of semicrystalline, conjugated polymer systems in recent years. However, progress has been hampered by the poor understanding of the interplay between the complex microstructure and measured charge mobility of these materials. The molecular weight~\cite{Singh2013,Brinkmann2007,Wu2010}, solution processing technique~\cite{Higashi2011}, crystallinity~\cite{Balko2013}, regioregularity~\cite{McMahon2011,Kohn2012,Sirringhaus1999} and coupling defects~\cite{Kohn2012}, all contribute to the microscopic morphology and, by extension, the transport properties such as the mobility. 

Many studies have been devoted to elucidating the fundamental phenomena underpinning the chief obstruction to satisfactory photocurrent collection; the trapping of charge-carriers in the amorphous grain boundaries separating crystalline domains~\cite{Lee, Jimison2009, Pandey2011, Hallam2009, Singh2013}, and its influence on charge transport. 
Efros and Shklovskii first applied the principles of percolation theory to low temperature hopping transport in amorphous materials~\cite{Efros}.
Their framework however, cannot be extended to higher temperature experimental systems and as such, several lattice models have been devised to describe charge-transport in disordered media.
The semi-empirical Gaussian disorder model was constructed from the results of Monte Carlo variable range hopping simulations on a 3D lattice~\cite{Pasveer2005,Bassler1993}. It has been applied to empirically fit the experimentally observed Poole-Frenkel type square-root dependence of the charge-carrier mobility on the electric field~\cite{Mozer2004, Mozer2005}, and has also been used in conjugated polymer simulations~\cite{Nelson}. Subsequent studies invoked the concept of a correlated Gaussian density of states (DOS) arising from charge-dipole interactions~\cite{Norikov1998,Parris2000}. 
%Percolation hopping theory was employed to introduce an effective temperature parameter to explain the field-effect mobility in small organic molecules ~\cite{Vissenberg1998,Jansson2008}.
%, while other works focused on rate-limited hopping from an effective transport energy ~\cite{Baranovskii}. 
%In parallel, several papers have characterised structural disorder within a paracrystallinity model as variations in the lattice spacings ~\cite{Rivney2011, Noriega2013, Poelking2013}.

In spite of all these efforts, it is still difficult to single out the crucial effects of macromolecular morphology on charge-transport in physical terms. The main problem is that all these models where originally derived to describe transport through molecular systems and do not contain the two coupling parameters required to describe the distinct mechanisms of interchain and intrachain hopping in conjugated polymers. The presence of these different transport mechanisms leads to multiscale mobility measurements where charges move faster over short lengthscales and slower over mesoscopic distances~\cite{NoriegaPNAS, Devizis1, Devizis2, Devizis3}. 
%which produce different mobility scales at different lengthscales which dictate mesoscopic transport through conjugated polymers.
This paper presents a methodology which suggests a possible macromolecular mechanism of charge-transport combining ad hoc experiments with a theoretical analysis which implements the role of polymer chain topology to link the macroscopic mobility with the chain-level topology. Within our model, the amorphous polymer structure leads to important qualitative physics in the field-dependence of the mobility due to the competition between charge-transport along the polymer backbone and activated variable-range hopping (VRH) from one chain to the other. 

\section{Materials and Methods}
\subsection{Experimental system}
Transport in regioregular poly(3-hexylthiophene) (RR-P3HT) was experimentally probed via Time-of-Flight (TOF) photocurrent measurements.  Both defect free and ``Single-defect'' RR-P3HT with one single tail-to-tail (TT) defect randomly located within a polymer chain were studied here. The details of the synthesis are given in~\cite{Kohn2012}. Small signal current mode Time-of-Flight mobility determinations were performed on capacitor-like samples prepared as described below fitted with electrically blocking contacts. The device was subjected to a voltage bias V$_0$ and optically excited by a strongly-absorbed, 532 nm wavelength pulse of a frequency doubled sub-nanosecond solid-state laser, with subsequent generation of a photocurrent decay. Electrical current transients for various V$_0$ were amplified, digitized and processed in order to extract the hole transit time ${\rm t_{tr}}$ across the device. Hole mobilities $\mu$ were thereby calculated according to the simple formula $\mu = D^2 /{\rm V_0 \, t_{tr}}$, where $D$ is the device thickness. This parameter was chosen to ensure that ${\rm t_{tr}}$ was greater than the time taken for the holes to reach quasiequilibrium, thus producing non-dispersive mobility measurements~\cite{Bassler1993}.

\subsection{Device preparation}
Device fabrication was performed in a glovebox environment (water and oxygen content $ <0.1$ ppm), where RR-P3HT films were formed by casting a drop of solution onto a transparent substrate and allowing the solvent to slowly evaporate. Typical active layer thickness varied between 2.5 $\mu$m and 3.5 $\mu$m (but reaching 10.1 $\mu$m for the lowest molecular weight). The substrates used were 12 mm wide, 1.1 mm thick soda lime square slides pre-coated with a longitudinal 7 mm wide, 120 nm thick indium tin oxide stripe. Prior to RR-P3HT solution casting, the substrates were sensitized with polyethylenimine, 80\% ethoxylated (PEIE) as described in~\cite{Zhou2012}, and after the deposition of the active layer, MoO$_3$--Ag top contacts were thermally evaporated onto the dry polymer film~\cite{Kyaw2013}. 

\subsection{Time-of-Flight measurements of electrical mobility}
Mobility scans were performed at several temperatures, since samples were held in a customised continuous flow, helium-cooled optical cryostat system. The intensity of the incident laser light was limited so that the integrated photogenerated charge was largely kept below $\rm 0.1 \times C \, V_0$, where C is the geometric capacitance of the device -- in order to ensure the uniformity of the internal electric field.

\section{Theoretical Framework}
\subsection{Model assumptions}
We begin by briefly listing all the assumptions implicit in the derivation of our mobility model, while further discussion of their applicability to the semicrystalline conjugated polymer P3HT is postponed until later in this paper. 

\subsubsection{Absence of bridging-chain percolation}
This theory is applicable to semicrystalline conjugated polymers in which the ordered domains are not connected by a complete percolating network of bridging chains, and therefore charge-carriers must hop from one chain to the next in the amorphous grain boundary regions in order to traverse through the material. This condition holds for samples in which the conjugation length of the polymer chains is comparable to the period length or in which the chains fold back into the original crystalline domain as opposed to bridging through to the next one.

\subsubsection{Amorphous domains control charge-transport}
Secondly, while significant overlap of the conductive $p_{z}$ orbitals in highly ordered crystalline domains facilitates fast charge-carrier transport, structural disorder in the amorphous grain boundary domains disrupts these rapid inter-chain hopping pathways and thus these regions act as the rate-limiting steps to the movement of charges through the material. In the absence of interconnecting bridging chains, the slow mobility of charge-carriers through these disordered regions effectively characterises the transport properties of the entire semicrystalline conjugated polymer sample. 

\subsubsection{Generalised VRH assumption for inter-chain hops}
The slow rate of inter-chain hopping in the disordered regions, together with weak electron-phonon coupling, allows one to model transport through such amorphous regions within a generalised variable range hopping model, with each hopping site representing one localized state. Here, faster conductive intra-chain transport acts as a perturbation which can be captured within a single corrective fitting factor, calibrated to experimental data.  

\subsubsection{Localization length}
The polymer chain can be split up into discrete hopping sites or localized states.%, each consisting of a straight segment of length equal to the persistence length.% ($\sim$ 5 monomer units).

\subsubsection{Random isotropic distribution of hopping sites}
It is assumed that polymer chain bending and twisting leads to the random orientation of the conductive $p_{z}$ orbitals  relative to one another in the amorphous regions. This allows one to cast inter-chain hopping in the disordered regions as a thermally activated process from one effectively isotropic localized state to another. 
A disordered region can thus be described by an isotropic homogeneous distribution of hopping sites in 3D space, with no correlation between structure and electric field direction. This is a reasonable first order approximation in the case where no experimental information about the polymer morphology in the amorphous grain boundary regions is at hand.  It follows that the distributions of accessible hopping sites around any two given hopping sites are identical provided that both sites have the same site energy.

\subsubsection{Distribution of site energies}
Seventhly, the distribution of site energies in the disordered regions is assumed to be Gaussian.

\subsubsection{Rule of fastest hopping rate} 
Finally, the model assumes that out of the complete set of possible hops, the charge-carrier will always choose the one with the fastest hopping rate. This condition is justified by the exponential factor in activated kinetic rate theory.

%2) Sufficiently disordered that amorphous region is the rate limiting region. Matthiessen's rule.
%2) Intra-chain hopping is faster than inter-chain hopping such that it can be treated as a perturbation to a discrete site variable range hopping treatment. Weak %electron-phonon coupling.
%3) Pz orbitals are oriented in a random fashion relative to one another in the amorphous grain boundaary regions. Approximate as spherical wavefunction.
%4) Polymer chain can be split into discrete hopping sites each consisting of one persistence length.
%3) Mean field approach, isotropic, homogeneous distribution of hopping sites in 3D space around every hopping site with no correlation between structure and field %direction. Completely disordered. Reasonable given that we have no information about the structure. Discussed further in Section 4. The distributions of hopping %sites around any two given hopping sites are identical provided that both sites have the same site energy U
%5) Gaussian density of states, one site one state.
%6) we assume that the charge-carrier always hops to the site with the smallest R. Reasonable given the expoential function.

\subsection{Input parameters}
Next, we specify all the model input parameters and their physical meaning. Firstly, there are three environmental parameters, namely the applied electric field strength, the charge concentration and the temperature, which can be varied to replicate the relevant application settings. Next, there are the four material parameters; the sample crystallinity (the fraction of the sample composed of ordered crystalline domains), the wavefunction localization length (effective radial extent of the conductive $p_{z}$ orbital), the material density and the attempt-to-escape frequency (averaged phonon frequency). Then, there are the site parameters which depend on how the discrete hopping sites are defined along the polymer chain. These are the site localization length along the chain, the Gaussian variance of the density of site energies and the Fermi energy.

Finally, the topological correction ratio $N_{corr}$ is defined as the number of hopping sites which the charge-carrier can hop to within a defined time frame during which intra-chain transport is active or permitted, divided by the number of sites which are accessible via bare inter-chain hopping in the absence of intra-chain transport. Note, that values for all of the above parameters are obtained either through direct measurement or from the literature, except for the final parameter $N_{corr}$ which is estimated by calibrating the model to our experimental data.  A schematic illustrating how all the model input parameters contribute to the derivation of the charge-carrier mobility is shown in Fig. 4 in Appendix A.
%electric field
%temperature
%Sample crystallinity
%Escape frequency
%Average density of sites
%Average spatial distance between discrete sites....Density of amorphous P§HT...Persistence length
%Gaussian variance of density of states
%Fermi energy
%Inverse wavefunction localization length
%Correction factor

\begin{figure}
\centering
\subfigure
{\includegraphics[width=8cm]{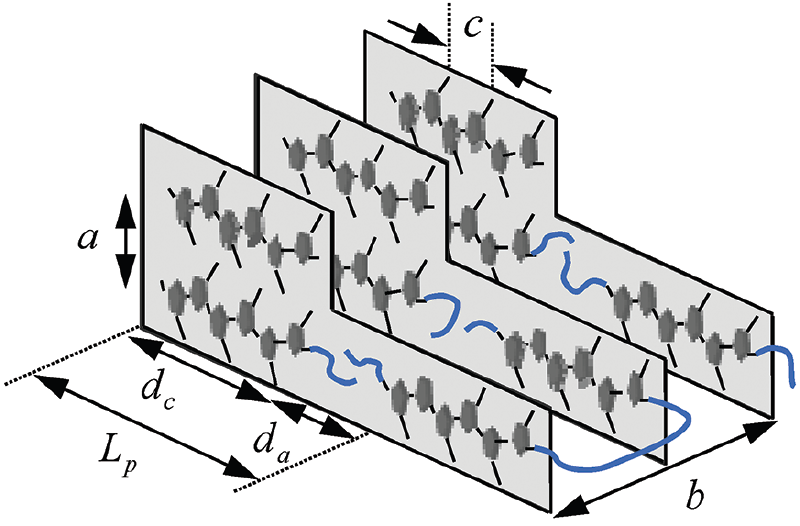}
\label{fig:model}}
%\begin{picture}(0,0)(0,0)
%\put(-230,10){(a)}
%\end{picture}
%\subfigure
%{\includegraphics[width=0.80\columnwidth]{fig1b.png}
%\label{fig:gr_max}}
%\begin{picture}(0,0)(0,0)
%\put(-230,10){(b)}
%\end{picture}
\caption{(Color online) Mesoscopic material structure and basic VRH assumptions.
Schematic of the structure of the typical conjugated polymer $P3HT$. Note key features such as the 2D lamella (aligned in the a-c plane), the $\pi-\pi$ stacking (in the $\vec{b}$ direction), side chains, the crystalline domain and the amorphous grain boundary region.}
\end{figure}

\subsection{Morphology}
As evidenced in Fig. 1, the molecular structure of semicrystalline P3HT used in solar energy devices comprises two regions; highly ordered crystalline domains and the extended, amorphous grain boundaries. In the absence of the interconnection of the former regions by a complete percolating network of conductive bridging polymer chains, the disorder in the latter regions strongly affects charge transport and acts as the rate-limiting component of the current-carrying pathway through the material. 

In contrast, significant $\pi-\pi$ conjugated-orbital overlap in the highly ordered crystalline domains results in charge delocalization along the $\pi-\pi$ stacking direction ($b$ axis)~\cite{Sirringhaus1999}. This facilitates rapid transport through the crystalline domain rendering its contribution negligible when we compute the overall mobility $\mu_{tot}$ of the material~\cite{Crossland2012, Jimison2009, Orton1980}. That is to say, according to Matthiessen's rule
$\frac{L_{p}}{\mu}=\frac{d_{c}}{\mu_{cryst}}+\frac{d_{a}}{\mu_{dis}}$, from which 
$\mu\simeq L_{p}\mu_{dis}/d_{a}=\mu_{dis}/(1-x)$ follows because $\mu_{cryst}\gg \mu_{dis}$. 
Here, $\mu_{cryst}$ is the mobility in the crystalline domains, $\mu_{dis}$ is the mobility in the amorphous grain boundary regions, $x=0.65$ (estimated according to~\cite{Kohn2012}) is the crystallinity of the polymer sample and all relevant length-scales are referenced in Fig. 1.

Hence, we approximate the overall mobility of the material with the mobility of the amorphous region on which we focus in the following.
Furthermore transport through these regions is slowest when the polymer chain does not bridge through to another crystalline domain, and thus the mobility through the amorphous region can be recast as the inter-chain hopping mobility, $\mu_{dis}=\mu_{inter}$ . The task of calculating the overall mobility is now reduced to estimating the \textit{effective} inter-chain hopping mobility $\mu_{inter}$ in the amorphous regions.
%, by properly accounting for the effect of fast transport along the chain (intra-chain transport), as discussed in the main article. 

The dimensionality of charge transport in the disordered grain boundary regions must be discussed before developing the model. Charge-transport is effectively two-dimensional in the crystalline region where the conducting $p_{z}$ or $\pi$ orbital is aligned parallel to the $b-c$ plane thus impeding transport along the $a$ axis (see Fig. 1). In contrast, by assuming the orientation of the $p_{z}$ orbital to be isotropic in the amorphous region, charge-transport becomes three-dimensional in the disordered region. Hence, our theoretical model as developed in the following is three-dimensional in space since it applies to the amorphous regions which effectively control the overall mobility of the sample. 
%In the following, we shall first introduce VRH theory for inter-chain hops on a disordered lattice, and we subsequently extend it to include the effect of $\pi-\pi$ conjugation and polymer chain topology.

There are two transport mechanisms which are active in the amorphous region: inter-chain and intra-chain transport. As the orientations of the conductive $\pi$ orbitals are uncorrelated in the disordered regions, angular isotropy allows one to characterise inter-chain transport as thermally-activated hopping from one hydrogen-like localized state to another on a different chain. 

Lifshitz's self-averaging assumption can then be applied to produce an wavelength-independent decay length, as averaging over all generated charge-carriers renders any statistical deviation from this  isotropic distribution negligible.

Note that due to the lack of experimental structural information about the polymer chain morphology in the amorphous regions, the distribution of inter-chain hopping sites is approximated to be homogeneous and isotropic everywhere. Also, along the chain, delocalization of unpaired electrons may be limited to discrete segments of the chain~\cite{Heeger_book} and electrons must thermally hop from one segment to the next in order to travel along the chain. 

At the same time, computer simulations of disordered P3HT have also estimated the wavefunction localization length to extend over about 5 monomers~\cite{Vukmirovi_2011, Vukmirovi_2013} along the polymer chain. 
It is thus reasonable to take 2.1 nm (equivalent to about $5$ monomer units), as the characteristic length scale for delocalized intra-chain transport in amorphous P3HT. 
%As the characteristic length scale for delocalized intra-chain transport it is meaningful to take the polymer persistence length (2.1nm, equivalent to $5-6$ monomer %units). 
Therefore, current-carrying pathways can be schematically described as a mixed sequence of phonon-assisted inter-chain hops and faster intra-chain transport. We first consider the inter-chain mobility $\mu_{inter}$ as it applies to standard disordered lattices, and we then extend it to account for the transport along the polymer chain. We neglect electron-phonon coupling and polaronic effects because they have been shown to be negligible in amorphous P3HT due to the large energy barriers involved~\cite{vukmirovic1,vukmirovic2}.

\subsection{Inter-chain transport}
The conduction of charge-carriers through disordered media is described by Mott's variable range hopping (VRH) model. Within this framework, the rate a charge-carrier hops from one localized site to another randomly distributed localized site scales as $\kappa \sim \exp (-2\alpha r - W)$~\cite{Mott}, where $\alpha$ is the inverse decay length of the hydrogen-like localized state wavefunction, $r$ is the absolute value of the distance vector between the two sites in 3D space, and $W=(V-U)/k_{B}T$ is the difference between the energy $V$ of the initial site and that of the final site, $U$. At high $T$, typically close to room temperature or above, thermal activation from the phonon bath dominates, such that only hops to the spatially closest nearby sites can optimize the hopping rate $\kappa \sim \exp[-(V-U)/k_{B}T]$, and the mobility displays Arrhenius dependence on temperature~\cite{Brenig}. At lower T, however, hops towards more distant sites for which the energy barrier is smaller or vanishing, become energetically favourable and one can define the variable range $R=2\alpha \xi + W$ in 4 dimensions (three in space, one in energy), such that $\kappa \sim \exp (-R)$.

Under an applied electric field $F$, upon assuming a random spatial distribution of sites and that the charge-carrier always hops to the site with the smallest range $\bar{R}_{nn}$,  one arrives at the following VRH form for the mobility~\cite{Apsley1975}
\begin{equation}
\mu_{dis}(U) = \frac{\nu_{ph}}{F}\bar{r}_{F}(U)\exp(-\bar{R}_{nn}(U)).
\end{equation}  
Here $\nu_{ph}$ is the average escape-attempt frequency, conventionally approximated with the average phonon frequency of the system and $\bar{r}_{F}$ is the spatially-averaged field-component of the hopping vector. Next, mean field theory stipulates that the distributions of accessible hopping sites around any two given hopping sites are identical provided that both sites have the same site energy $U$. This condition allows one to average Eq. 1 over all possible occupied site energies $U$ to recover the macroscopic inter-chain hopping mobility in the amorphous regions.

In the limit of weak electron-phonon coupling, the variable range to be optimized is given by the standard Miller-Abrahams rate~\cite{Ambegaokar}  
\begin{equation}
R = \begin{cases}
2\alpha r + \frac{V - U - Fqr\cos\theta}{k_{B}T} & \text{if $V - U - Fqr\cos\theta > 0$} \\
2\alpha r & \text{if $V - U - Fqr\cos\theta < 0$},
\end{cases}
\end{equation}
where $\theta$ is the angle between the hopping direction and the field.

Under the assumption that the DOS is constant near the Fermi level and the field is low, the famous Mott's law $\mu_{dis}\sim \exp(-1/T^{1/4})$ can be readily derived by simply optimizing the above expression for $\kappa$ with respect to $R$ over the homogeneous isotropic distribution of sites. In a more rigorous formulation~\cite{Apsley1975}, if $N$ is the total number of sites available within a contour $R$, the smallest nearest neighbour range in four dimensional range-space is given by
$\bar{R}_{nn}(U)=\int_{0}^{\infty}\exp[-N(R,U)]dR$. Hence, $N(R,U)$ is the key quantity for the mobility calculation, and depends on the electronic DOS, on the spatial distribution of sites (i.e. the packing density and size of the monomers in our case) and on the applied field $F$.

Mott's original VRH theory is valid at low T where the constant-DOS ansatz is reasonably valid, and at high charge-carrier densities. Our more general framework with a non-constant, Gaussian DOS, instead, is valid over a much broader range of temperatures up to and including the thermal activated regime, and for low charge-carrier densities as well. 
Our interest here is to implement the role of the polymer chain topology within this extended VRH framework. 
%The key difference with respect to monoatomic semiconductors is that charges in polymers can explore nearby sites via quasi-delocalized intra-chain conduction prior to hopping to another polymer chain. 
In order to account for the interplay between \textit{intra-chain conduction} and \textit{inter-chain hopping} we shall introduce new quantities specific for the polymer topology.
Furthermore, our model is more general than standard VRH theory because we implement a Gaussian DOS, which is a good model for amorphous polymers as shown by many studies in the past~\cite{Bassler1993}.

\subsection{Intra-chain transport}
The first new quantity is the number of sites $N_{inter}$ on different chains which the charge can visit within a time-scale $\tau_{1}(R_{1}) \sim P(R_{1})^{-1}=\exp (R_{1})$ where $R_{1}$ is a contour in four-dimensional range-space enclosing a test site 1, as schematically depicted in Fig. 2(a). Next we consider the role of the polymer-chain topology allowing for faster intra-chain transport events which may precede the slower inter-chain hop. Clearly, if the charge can travel along the chain backbone as a partly-delocalized wavefunction, then the number of sites accessible (energetically favourable) on nearby chains within time $\tau_{1}$, increases. 

\begin{figure}
\centering
\begin{subfigure}
{\includegraphics[width=9.75cm]{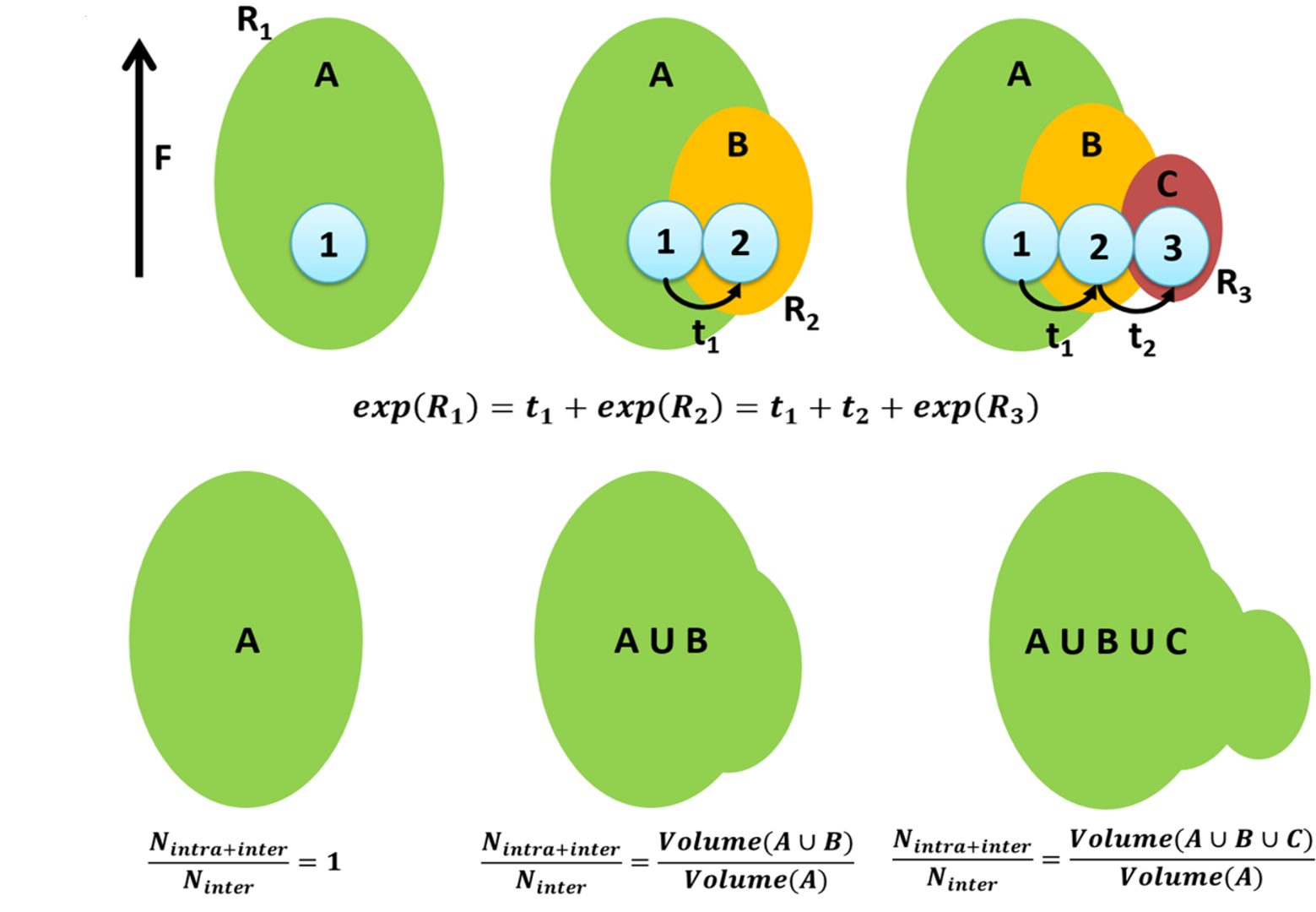}
\label{fig:model}}
\begin{picture}(0,0)(0,0)
\put(-295,0){(a)}
\end{picture}
\end{subfigure}
%\subfigure
%{\includegraphics[width=0.80\columnwidth]{fig2b}
%\label{fig:gr_max}}
%\begin{picture}(0,0)(0,0)
%\put(-220,10){(b)}
%\end{picture}
\begin{subfigure}
{\includegraphics[width=8cm]{fig2b}
\label{fig:gr_max}}
\begin{picture}(0,0)(0,0)
\put(-268,10){(b)}
\end{picture}
\end{subfigure}
\caption{(a): Schematic of the effect chain morphology has on the shape of constant hopping rate contours in four dimensional range space. (b) Schematic of topology-controlled inter-chain charge transport in an electric field. The charge-carrier may hop directly onto the other chain (solid arrow) or exploit chain morphology to reach it indirectly via multiple sequential hops involving fast intra-chain conductive transport (dashed arrows).}
\end{figure}

The mechanism is illustrated in Fig. 2(a). The charge-carrier is situated on the initial site 1 as shown on the left-hand side. In the presence of an electric field, the contour of four-dimensional range space the charge carrier can visit in time less $\exp(R_{1})$ is highlighted by the egg-shaped region $A$ (it would be spherical in the limit $F=0$). Here, $\tau_{1}=\exp(R_{1})$ indicates the maximum time-scale to reach any other site via a single inter-chain hopping event that starts from site 1. We define $t_{1}$ as the time it takes from site 1 to reach the nearest site 2 along the same chain, and we let the contour $B$ contain the sites on other chains the charge can hop to from site 2 in time $\exp(R_{2 })<\exp(R_{ 1})$. 

It follows that the total volume in range-space encompassed by the $R_{1}$ contour increases from $V(A)$ to $V(A \cup B)$ thanks to the contribution of conduction along the chain to the nearest-neighbour monomer 2. The volume $B$ is smaller than $A$ because it takes into account that 1D transport along the chain may be finite due to the presence of defects, chain folds, kinks etc., and also because the increased number of favourable sites which are made accessible for hopping makes the smallest range $\bar{R}_{nn}$ effectively decrease. In other words, the charge does not need to hop to a distant site within the original contour $V(A)$ but it can land onto closer sites within $V(B)$ or $V(C)$. Overall the increased density of accessible sites, thanks to intra-chain fast transport, leads to a shorter optimal range $\bar{R}_{nn}$ and, according to Eq.(1), boosts the mobility to larger values (compared to a lattice of isolated point-sites with no polymer chains).

We now introduce the parameter $N_{intra}$ to denote the number of extra sites in range-space which become available thanks to this effect, and use $N_{inter}$ to denote the standard number of available hopping sites on other molecules within VRH. $N_{intra}$ can be formally defined by iterating the mechanism to the next-nearest monomers down the initial chain. If we add another monomer 3 as displayed on the right of Fig. 2(a), assuming $t_{3}=t_{1} + t_{2} < \exp(R_{1})$ (where $t_{2}$ is the time to reach site 3 from site 2), our value of $N_{inter+intra}/N_{inter}$ will increase further as schematically shown. This process can be iterated until either the end of the chain is reached, or $\sum_{i=1}^n t_{i} > \exp(R_{1})$. 

In a fully disordered media, we assume that the density of available hopping sites is always spatially homogeneous in the vicinities of every starting site, and thus this effect can be accounted for within VRH theory simply by multiplying $N_{inter}$ by the correction parameter $N_{corr}=N_{inter+intra}/N_{inter} > 1$ where $N_{inter+intra}/N_{inter} = V(A \cup B \cup C \cup ...) / V(A)$. With this correction in VRH theory, we can analyse and disentangle the competing effects of inter-chain and intra-chain transport. 
%Also, it is possible to study the effect of positional disorder and coupling defects which may affect intra-chain transport and hinder the number of accessible paths to the next chain.

\subsection{Application to experimental data} 
Our calculation of the mobility (reported in full detail in Appendix A) comprises the following steps. First, the number of available sites for inter-chain hopes $N_{inter}$ is calculated as a function of the Gaussian DOS~\cite{Bassler1993}, of the field $F$ and for a homogeneous isotropic distributions of monomers, according to standard VRH hopping theory in the quantitative formulation of Ref.~\cite{Apsley1975}. This estimate of $N_{inter}$ does not differ from the estimate of the number of hopping sites $N$ in monoatomic amorphous semiconductors, where now the monomers (in lieu of the atoms) play the role of the basic unit for transport. We then introduce the topological correction parameter $N_{corr}=N_{inter+intra}/N_{inter}$ such that the total number of available sites, accounting for both inter-chain and intra-chain transport, follows as $N_{tot}=N_{inter}N_{corr}$. The limit of monoatomic amorphous solids, where there is no effect of chain topology, is recovered for $N_{corr}=1$. 
%Furthermore, this new structural parameter may account for the non-trivial morphology of polymer chains in glassy polymer regions, where folds, kinks and other "defects" can result in lower values of $N_{corr}$ as a consequence of lower values of $N_{intra}$ because such defects act as to slow down charge transport along the chain. 
The smallest range $\bar{R}_{nn}$ which enters the definition of mobility, is thus calculated using the total number of available sites defined in this way as $\bar{R}_{nn}=\int_{0}^{\infty}\exp[-N_{tot}(R)]dR$.

In this calculation, $N_{corr}$ is the only parameter which is extracted from the quantitative match with the experimental mobility data. All other parameters which enter in the model are either specified by the experimental protocol or taken from the literature on P3HT as detailed in Appendix A.
From this fitting, deeper insights into the role of polymer topology on charge transport can be obtained because the evolution of the topological correction factor $N_{corr}$ as a function of temperature and electric field, encodes information about the interplay between inter-chain and intra-chain transport mechanisms under different applied environmental settings.
%From this fitting, deeper insights into the role of polymer topology on charge transport can be obtained because the topological parameter $N_{corr}$ encodes the %information about the interplay between inter-chain and intra-chain transport mechanisms.

\begin{figure}
\centering
\subfigure
{\includegraphics[width=9cm]{Fig3a}
\label{fig:model}}
\begin{picture}(0,0)(0,0)
\put(-260,10){(a)}
\end{picture}
\subfigure
{\includegraphics[width=9cm]{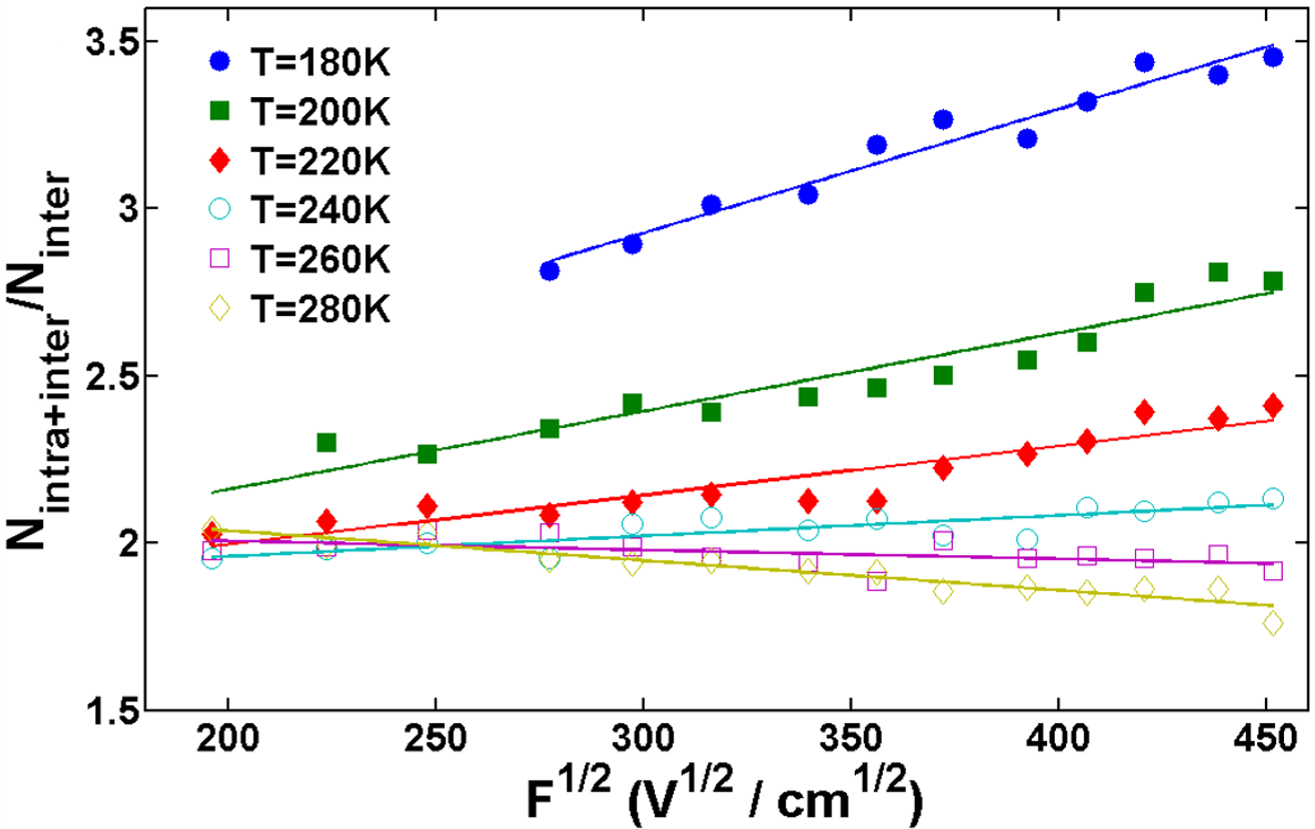}
\label{fig:gr_max}}
\begin{picture}(0,0)(0,0)
\put(-260,10){(b)}
\end{picture}
\caption{(Color online) Comparison between experimental data of mobility measured by TOF in a semicrystalline P3HT sample as a function of the applied electric field and theoretical fits with the adjustable topological parameter $N_{intra+inter}/N_{inter}$.
%Mozer 2005 shows mobility measurements are non-dispersive above 180. Carlo measured them to be non-dispersive anyway
(a) Non-dispersive experimental mobility measurements (symbols) as a function of the electric field for a sample with one TT defect per chain and molecular weight given by 11.8 kg / mol. Solid lines are theory calculations using the structural parameter $N_{intra+inter}/N_{inter}$ as an adjustable parameter, see text.  (b) Plot of the fitted $N_{intra+inter}/N_{inter}$ as a function of the electric field for a sample with one TT defect per chain and molecular weight given by 11.8 kg / mol.
The charge concentration is measured as $7.81\times 10^{13} cm^{-3}$, the period length $L_{p}$ as $16.7$ nm, the average polymer chain conjugation length as $16$ nm and the crystallinity $x$ as 0.65. The inverse wavefunction localization length $\alpha$ (spatial extent of the $p_{z}$ orbital) was computed as $7.43\times 10^{9} m^{-1}$, the wavefunction localization length along the chain $l$ as $2.1 nm$, the variance in the Gaussian density of states $\sigma$ as 60 $meV$, the material density as $1.05 g/cm^{3}$ and the attempt-to-escape frequency $v_{ph}$ as $4.50 \times 10^{13} s^{-1}$. The Fermi energy $\epsilon_{F}$ was found to vary between $-0.46 eV$ and $-0.36 eV$ over the measured temperature range. Details of the calculation of all the model input parameters are included in Appendix A. 
}
\end{figure}

\section{Results and Discussion}
The fitting of the mobility data using our model as a function of the field is shown in Fig. 3(a), while the structural parameter $N_{corr}$ obtained from the fitting is plotted in Fig. 3(b) for the same conditions. 
As expected, the outcome that $N_{corr}>1$ for all curves, is physically meaningful: fast intra-chain transport enlarges the number of easily accessible hopping sites on nearby polymer chains within a given range, and reduces the optimum range $\bar{R}_{nn}$. 
Another important insight that we obtain from this analysis is that $N_{corr}$ consistently decreases with increasing $T$. 
This is again a meaningful outcome: inter-chain transport is slower and thermally-activated, by definition, whereas intra-chain transport is faster and less sensitive to $T$ because it is boosted by delocalized transport along the chain. For these reasons, upon increasing $T$, $N_{inter}$ increases more significantly than $N_{intra}$ which causes the ratio $N_{corr}=N_{inter+intra}/N_{inter}$ to decrease. 

We conclude that the temperature dependence of the slope, which is observed in all the measured data, is stronger in samples where obstructive TT defects restrict intra-chain transport. This boosts the impact that the restricted intra-chain transport with greater electric fields has on the overall transport through the disordered regions,  over the measured temperature range. The clearest illustration of this slope inversion is provided by the 1TT 11.8 kg/mol sample in Fig. 3, where the reader can observe that $N_{corr}$ is significantly decreasing as a function of increasing electric field at the highest recorded temperatures.
Note however, that neither the polydispersity nor the presence, or lack thereof, of conjugation defects alters the qualitative trends observed in Fig. 3 or any of the results presented in this paper.
A more detailed analysis of the influence the molecular weight and conjugation defects have on the transport properties of semicrystalline conjugated polymers will be contained within a follow-up paper currently under preparation.

Our theory can be used to explain the Poole-Frenkel qualitative dependence $\log\mu \propto \sqrt{F}$ for semicrystalline conjugated polymers.
At low T, thermally-activated inter-chain hops are slow and long hops to low-energy sites are possible which would cause the optimum range $\bar{R}_{nn}$ to be large. A crucial contribution to lowering the optimum range $\bar{R}_{nn}$ by increasing the number of easily accessible sites within a shorter range comes from intra-chain transport. At $T=180K$ this is reflected in high values of the topological parameter $N_{corr}=N_{inter+intra}/N_{inter}\gtrsim 3$, which means that the total number of available sites within a given range is about three times as large compared to the value one would have without the effect of intra-chain transport. 
Hence, at low T, inter-chain transport is controlling (it is the slowest mode), and the mobility increases markedly upon increasing $F$ because only down-field inter-chain hops are allowed (hops with a component against the field are energetically forbidden~\cite{Ambegaokar}). 

Increasing T has two main effects. On one hand, inter-chain hops, having an increasingly important thermally-activated contribution, become faster and the number of sites reachable with inter-chain hops within a given range increases. Clearly, this means that $N_{inter}$ increases with T, which in turn causes $N_{corr}$ to decrease with T: this is what we observe in Fig. 3(b). At high temperatures, this speed-up means that the intra-chain transport mode is no longer "instantaneous" compared to inter-chain hopping. In general, intra-chain transport, unlike inter-chain hopping, can also happen up-field, i.e. with a component opposite to the field. 

This situation is schematically depicted in Fig. 2(b). The pathway involving intra-chain transport with an up-field component (dashed line) can be more favourable than the single down-field inter-chain hop (solid line) at low fields. At high T where inter-chain hopping has become faster, however, pathways like the dashed one in Fig. 2(b) are no longer favourable at high $F$. 

This effect of intra-chain transport slowing down with increasing field is, in fact, always there. The field-dependent time-scale of intra-chain transport is given by 
$\tau_{intra} \propto\frac{1}{\pi}\int_0^\pi \exp(\frac{Fql\cos\theta}{k_{B}T}) 
\propto I_{0}(Fql/k_{B}T)$, where $I_{0}(x)$ is the modified Bessel function of the first kind, $l$ is the localization length of the hopping sites along the polymer, and $\theta$ the angle between the field and the pathway. Since $I_{0}(x)$ is an even and monotonic \textit{increasing} function of $F$, it is clear that the intra-chain transport slows down upon increasing $F$ due to the pathways with up-field components becoming unfavourable. 
At low T this effect is also there, of course, but since the inter-chain hopping is controlling in that regime, and inter-chain hopping can only become faster with increasing the field, the inter-chain effect determines the field-dependence. 

%Finally, in Fig.4 we plotted the mobility as a function of T for different values of applied field. In the Arrhenius plot, data points would lie on straight lines if the transport was controlled by thermal activation only. It is evident that the data points do not lie on perfectly straight lines, which means that thermal-activation is not the only mechanism. Especially at low T the data points depart significantly from a straight line in the log-linear plot, which confirms that VRH is the right framework to describe our data. In fact, we also fitted the T-dependence with Mott's law and found an equally good fitting ($R^{2}=0.994$), although not perfect because the assumption of constant DOS leading to Mott's law may not be tenable and in fact we used a fully calibrated Gaussian DOS for our quantitative comparison in Fig.2.  

One of the main assumptions of this model is that in the absence of experimental structural information about the polymer chain morphology in the amorphous regions, the distribution of inter-chain hopping sites is cast as homogeneous and isotropic everywhere. Relaxing this condition to the statement that the chains remain preferentially in or close to the plane of the film, as they do in the crystalline regions, does not alter the fact that chain bending and twisting is expected to significantly disrupt the $\pi-\pi$ stacking which enables rapid directional inter-chain transport in the crystalline regions. As the ToF experimental technique typically applies the field along the alkyl stacking direction, the structural disorder disrupts perfectly aligned inter-chain hopping in the pi-pi stacking direction, orthogonal to the plane of the film (a-c plane) and perpendicular to the electric field (along the $a$ axis). Therefore, possible inter-chain hops in the disordered regions are always either hindered or enhanced by the applied electric field. The same argument can also be applied to intra-chain hopping, which occurs along the $c$ axis in the ordered domains, perpendicular to the direction of the electric field. Any model in which all hopping rates in the amorphous regions, both intra-chain and inter-chain, are dependent upon the applied field intensity, preserves the qualitative trends observed in Fig. 3, making them a universal signature of transport through disordered conjugated polymer domains.

%One of the main assumptions of this model is that in the absence of experimental structural information about the polymer chain morphology in the amorphous %regions, the distribution of inter-chain hopping sites is cast as homogeneous and isotropic everywhere. Relaxing this condition to the statement that the chains %remain preferentially in or close to the plane of the film, as they do in the crystalline regions, does not alter the fact that chain bending and twisting is expected to %significantly disrupt the $\pi-\pi$ stacking which enables rapid directional inter-chain transport in the crystalline regions. As the ToF experimental technique typically %applies the field along the alkyl stacking direction, the structural disorder disrupting perfectly aligned inter-chain hopping in the $\pi-\pi$ stacking direction, which is %orthogonal to the plane of the film and perpendicular to the electric field, ensures that each possible inter-chain hop is either hindered or enhanced by the applied %electric field. The same argument can also be applied to intra-chain hopping, constrained to the plane of the film and perpendicular to the electric field. This %invariance preserves the qualitative trends observed in Fig. 3, making them a universal signature of transport through disordered conjugated polymer domains.

This theoretical framework describes the mobility of charge-carriers through semicrystalline conjugated polymers, where a combination of chain folds and chain ends forces the charge-carriers to hop from one chain to the next in order to traverse through the rate-limiting amorphous grain boundary regions. This condition is clearly satisfied by low molecular weights samples (such as that displayed in Fig. 3) where the conjugation length (16 nm) is too short compared to the period length $L_{p}$ (16.7 nm), for it to bridge through to the next crystalline domain. Here, the slow rate of inter-chain relative to intra-chain hopping allows us to include transport along the chain as a perturbation to the generalised variable range hopping model. 

At high molecular weights, it may be possible for the chains to fold, end or bridge through to the next crystalline domain. In such regimes, we expect our model to remain applicable albeit with an unknown scaling factor to account for the fraction of chains entering the amorphous regions which bridge through to another crystalline domain. It is only in structures where this fraction is sufficiently high that inter-chain hopping in the disordered regions becomes insignificant and our model is no longer directly applicable. Such polymer morphologies are fundamentally different and should be treated by the orthogonal frameworks derived by Salleo, Spakowitz and co-workers~\cite{NoriegaPNAS, Mollinger, Noriega2013}. There, the crystalline regions are interconnected by a complete percolating network of bridging chains and thus, intra-chain and inter-chain hopping in both amorphous and crystalline regions must all be treated at the same level.

%This theoretical framework describes the mobility of charge-carriers through semicrystalline conjugated polymers, where a combination of chain folds and chain ends %forces the charge-carriers to hop from one chain to the next in order to traverse through the rate-limiting amorphous grain boundary regions. Here, the slow rate %of interchain relative to intachain hopping allows us to include transport along the chain as a perturbation to the generalised variable range hopping model. This %polymer morphology is fundamentally different to the systems investigated by Salleo, Spakowitz and co-workers~\cite{NoriegaPNAS, Mollinger, Noriega2013}. %There, the crystalline regions are interconnected by a complete percolating network of bridging chains and thus, intrachain and interchain hopping in both %amorphous and crystalline regions must all be treated at the same level.

Polydispersity has no effect on the results as long as a complete percolating conductive network of polymer bridging chains remains absent from the sample. Our model does not require a defined value for the polymer chain length and remains valid and insensitive to the polydispersity index as long as we are satisfied that charge-carriers must hop from one chain to the next in order to traverse through the rate-limiting amorphous grain boundary regions.

The salient question is for what polymer chain length distributions does this condition hold. Bridging chains become possible when the ratio of chain conjugation length to the period length approaches two. Therefore, identifying the fraction of polymer chains in the sample which exceed this threshold length scale requires information about the average and the width of the conjugation length distribution, the latter of which is intimately linked with the polydispersity index. After accomplished this task, the problem of distinguishing between a polymer chain which bridges through to another crystalline domain, and one which folds back into the original ordered domain, must still be solved. This question remains open and will form the basis of further investigations in the future.   

The authors do not claim that this theoretical framework produces a predictive calculation of the charge mobility, rather it constitutes a district departure from semi-empirical closed form VRH mobility models in the literature, where often a large number of constant fitting parameters are applied will little accompanying insight into the microscopic physical mechanisms behind them. 
%By applying the topological correction factor $N_{corr}$, the macroscopic mobility of a sample can be directly linked to the competing microscopic mechanisms of %inter-chain and intra-chain transport, to analyse the extent to which temperature, electric field strength, structural disorder, coupling defects and molecular weight %all contribute to the overall transport properties of the material. Only an analysis of the first two is presented in this paper, with the remaining results forming the %basis of another paper currently under preparation. 
Instead, the unique feature of our model is the capture of the balance between intra-chain and inter-chain hopping within a novel phenomenological factor, which allows us, for the first time, to directly map the macroscopic mobility to both microscopic transport mechanisms and analyse the extent to which temperature, electric field strength, structural disorder, coupling defects and molecular weight all contribute to the overall transport properties of the sample. In this paper, only an analysis of the first two is presented while the remaining results are being reported on in another paper currently under preparation. 

\section{Conclusions}
We presented experiments and a microscopic model of charge transport in semicrystalline P3HT, a paradigmatic conjugated-polymer material widely used for solar energy devices. While the mobility increases strongly with the field at low T, the trend becomes much flatter upon increasing T until the mobility eventually slightly decreases with the field at room temperature. This effect, of paramount importance for technological applications, can be interpreted with our proposed model which implements the role of polymer topology in the VRH hopping theory, as a consequence of the competition between intra-chain transport along the polymer (due to wavefunction delocalization) and phonon-assisted inter-chain hops from one polymer to the other. Our modelling approach can be generalised to experimental data derived from other semicrystalline conjugated polymer systems, to provide a wealth of new microscopic information about the role of polymer chain topology and point-defects, and it may open up new avenues for the characterization and optimization of charge-carrier transport in conjugated polymers.

\begin{appendix}
\section{Appendix A: Derivation of the chain-assisted VRH conduction theory}
\subsection{General framework}
In this appendix we present the complete mathematical derivation of the chain-assisted variable range hopping mobility $\mu_{inter}$ of charge-carriers in amorphous, conjugated polymers, leading to the quantitative evaluation of Eq. (1) in the main article. As discussed in the main article, Matthiessen's rule dictates that the macroscopic mobility $\mu_{tot}$ is determined by the mobility $\mu_{dis}$ through the disordered grain-boundary regions. In the following, we shall calculate this quantity by first introducing VRH theory for inter-chain hops on a disordered lattice, and subsequently extending it to include the effect of $\pi-\pi$ conjugation and polymer chain topology. A flowchart depicting the structure of this calculation is shown in Fig. 4.

\begin{figure*}
\centering
\subfigure
{\includegraphics[width=16cm]{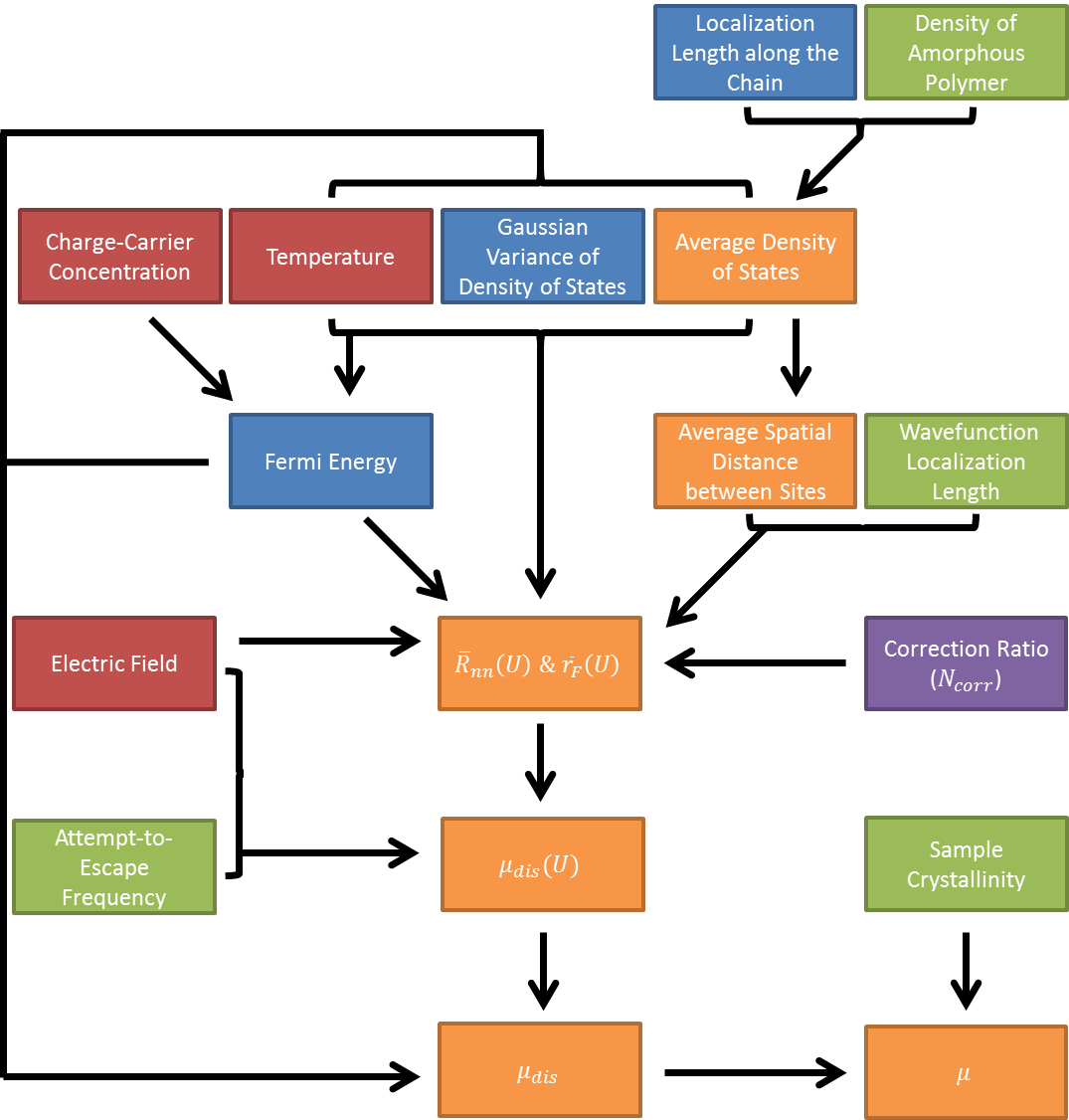}
\label{fig:model}}
\caption{(Color online) Flowchart of the structure of the mathematical mobility model. Environmental, material, site and fitted input parameters are shown in red, green, blue and purple respectively. Derived quantities are shown in orange.}
\end{figure*}

A variable range hopping framework is implemented to describe the dynamics of inter-chain hopping in disordered, organic semiconductors in an externally imposed electric field $F$. As outlined in Eq. (1) of the main article, the mobility $\mu$ of charge-carriers in such systems is~\cite{Apsley1975}
\begin{equation}
\mu_{dis}= \frac{v_{ph}}{F}\bar{r}_{F}\exp(-\bar{R}_{nn}).
\end{equation} 
The various contributions entering in this expression are estimated according to the procedure reported below.
The escape-attempt prefactor $\nu_{ph}$ is taken to be equal to the average phonon frequency in the system, $\nu_{ph}=4.50 \times 10^{13} s^{-1}$~\cite{Liu2011}.
$\bar{R}_{nn}$ is the dimensionless range or activation energy of the fastest available inter-chain hop, to be evaluated quantitatively in the next subsection. It is a function of both the energies of the site we are hopping to and from, and of the number of available sites for inter-chain hops including the contribution of intra-chain fast transport, $N_{corr}\cdot N_{inter}=N_{intra}+N_{inter}$. It is also a function of the 
the electric field $F$ and of the average spatial (radial) distance between two sites, $d$. The latter is calculated from the known density of amorphous P3HT ($1.05 g/cm^{3}$~\cite{Kohn2013}) and taking $2.1~nm$~\cite{Vukmirovi_2011, Vukmirovi_2013} as the characteristic linear size of a site. Using the linear size of a monomer ($0.385~nm$), corresponding to rigid segments or sites of $\sim 5$ monomer units, we estimate the density of sites as $n_{s}=6.97\times 10^{26}~m^{-3}$. From this, the spatial inter-site separation follows as $d\simeq n_{s}^{-1/3}=1.12~nm$. 

\subsection{Evaluation of $\bar{R}_{nn}$ and $N_{inter}$}
In strongly disordered regions the density of states $g(\epsilon)$ is well approximated by a Gaussian while the site occupation probabilities is governed by Fermi-Dirac statistics. We write the energy density distribution as $g(\epsilon)$ and the Fermi-Dirac distribution as $f(\epsilon,\epsilon_{F})$, where $\epsilon_{F}$ denotes the Fermi energy:
\begin{equation}
\begin{split}
g(\varepsilon)&=\frac{n_{s}}{\sigma\sqrt{2\pi}}\exp\bigg[-\frac{1}{2}\bigg(\frac{\varepsilon}{\sigma}\bigg)^{2}\bigg], \\
& -\infty <\varepsilon < \infty ;\\
f(\varepsilon, \varepsilon_{F})&=\frac{1}{1+\exp\bigg(\frac{\varepsilon-\varepsilon_{F}}{k_{B}T}\bigg)}.
\end{split}
\end{equation}
Here $n_{s}=6.97\times 10^{26}~m^{-3}$ is the density of hopping sites, while $\sigma=60~meV$ is the normal distribution variance estimated from an independent Baessler fit to experimental data using the method of Ref.~\cite{Bassler1993}. The Fermi energy $\epsilon_{F}$ is calculated from the following expression for the relative charge-carrier concentration $c$,
\begin{equation}
c=\frac{n_{c}}{n_{s}}=\frac{1}{n_{s}}\int_{-\infty}^{\infty} \! g(\varepsilon)f(\varepsilon , \varepsilon_{F}) \, \mathrm{d} \varepsilon
\end{equation}
where $n_{c}=7.81\times 10^{13}~cm^{-3}$ is the density of charge-carriers.

It is now possible to calculate $\bar{R}_{nn}$, the effective range of the fastest available hop. The derivation begins by ranking all unoccupied sites in order of increasing range $R$ in four dimensional (three spatial coordinates and the site energy) range space. The number of unoccupied sites which can be reached with hopping times varying between $0$ and $\exp(R)$ is denoted by $N(R)$. This quantity is exactly equivalent to $N_{inter}$ which is the quantity used throughout in the main article, $N(R)\equiv N_{inter}(R)$. Here we use $N(R)$ mainly as short-hand notation. 
The total number of empty sites with hopping times between $\exp(R)$ and $\exp(R+dR)$ is given by $\Delta N(R) dR$ where
\begin{equation}
\Delta N(R) = \frac{\partial N(R)}{\partial R}
\end{equation}
The next step is to evaluate the probability $P(R_{1})$ that the first nearest neighbour in this four dimensional space, corresponding to the fastest available hop, has a hopping time of $\exp(R_{1})$. This condition can be expressed mathematically as $P(R_{1})=\Gamma(R_{1})\Lambda(R_{1})$, where $\Gamma(R_{1})$ is the probability that there are no available sites with a hopping time of less than $\exp(R_{1})$ and $\Lambda(R_{1})$ is the probability that there exists exactly one empty site with a hopping time between $\exp(R_{1})$ and $\exp(R_{1}+dR_{1})$. Hence, it can be shown~\cite{Apsley1975} that the two distributions are binomial:
\begin{equation}
\begin{split}
\Gamma(R_{1})&=\bigg(1-\frac{N(R_{1})}{M}\bigg)^{M} \\
\Lambda(R_{1})&=\frac{M!(\Delta N(R_{1})dR_{1}/M)}{(M-1)!(1)!}\bigg(1-\frac{\Delta N(R_{1})dR_{1}}{M}\bigg)^{M-1}
\end{split}
\end{equation}
where $M$ is the total number of empty sites in the system. It follows that for large $M$,
\begin{equation}
P(R_{1})\simeq\frac{\partial N(R_{1})}{\partial R_{1}}\exp[-N(R_{1})].
\end{equation}
Hence the average range of inter-chain hops $\bar{R}_{nn}$ in four-dimensional range-space is
\begin{equation}
\bar{R}_{nn}=\int  \limits_0^\infty \, R_{1} P(R_{1}) \, \mathrm{d}R_{1}.
\end{equation}
Integrating by parts we retrieve,
\begin{equation}
\begin{split}
\bar{R}_{nn}&=\int \limits_{0}^{\infty} \, R_{1}\frac{\partial N(R_{1})}{\partial R_{1}}\exp[-N(R_{1})] \, \mathrm{d}R_{1} \\
&= \int \limits_{0}^{\infty} \, \exp[-N(R_{1})] \, \mathrm{d}R_{1}. \\
\end{split}
\end{equation}
In the second equality of Eq.(A8) we used the fact that the other term arising in the integration by parts (of the first integral) vanishes under the integration limits $R_{1}\rightarrow\infty$ and $R_{1}\rightarrow 0$. 
One should also note that $N(R_{1})$ is a function also of the energy $U$ of the site we are hopping from, of the temperature and of the electric field, as will be specified below. 

The final step required to derive the inter-chain hopping rate is to count the number of unoccupied sites $N(R_{1})$ with hopping times less than $\exp(R_{1})$. We begin by defining the two criteria a charge-carrier must fulfil in order to travel from one site to another. Firstly, it must tunnel through the dielectric and, secondly, thermal fluctuations (from the phonon bath) must provide enough energy to surmount the energetic barrier. As is common practice, we describe the dielectric as a vacuum. It follows that the range $R$ is a superposition of both transport mechanisms
\begin{equation}
R = \begin{cases}
2\alpha r + \frac{V - U - Fqr\cos\theta}{k_{B}T} & \text{if $V - U - Fqr\cos\theta > 0$} \\
2\alpha r & \text{if $V - U - Fqr\cos\theta < 0$}
\end{cases}
\end{equation}
with $V$ and $U$, the energies of the site we are hopping to, and from, respectively. $\alpha=7.43\times 10^{9}~m^{-1}$, is the inverse (hydrogen-like) wavefunction localization length (characterising the spatial extent of the wavefunction overlap), while  $r$ and $\theta$ denote the modulus of the hopping distance and the angle between the hopping vector and the electric field. Note that $\alpha$ is defined as the angle-averaged inverse localization length for inter-chain hops (not to be confused with localization length along the chain). The first term in Eq. (A9) is thus an effective tunnelling barrier-crossing term while the second term is the potential barrier for phonon-assisted hopping. We can rewrite the range in terms of dimensionless parameters
\begin{equation}
R = \begin{cases}
\bar{r}(\bar{\alpha}+\beta\cos\theta) + V' - U' & \text{if $V' - U' - \bar{r}\beta\cos\theta > 0$} \\
\bar{r}\bar{\alpha} & \text{if $V' - U' - \bar{r}\beta\cos\theta < 0$}
\end{cases}
\end{equation}
where $\beta=(Fed)/k_{B}T$, $V'=V/k_{B}T$, $U'=U/k_{B}T$, $\bar{\alpha}=2\alpha d$ and $\bar{r}=r/d$. Thus, we can write the number of empty sites with hopping times less than $\exp(R_{1})$, in spherical coordinates, as
\begin{equation}
\begin{split}
N &= k_{B}Td^{3}\int \limits_{0}^{R_{1}/\bar{\alpha}} \int \limits_{0}^{\pi} \int \limits_{-\infty}^{\rho} \, g(V')[1-f(V',\epsilon_{F})] \bar{r}^{2}\sin\theta \, \mathrm{d}V' \mathrm{d}\theta \mathrm{d}\bar{r}\\
\rho & = R_{1}+U'-\bar{r}(\bar{\alpha}+\beta\cos\theta).
\end{split}
\end{equation}
Note that the factor $k_{B}Td^{3}$ arises from normalization of the four-dimensional integral in range space (as usual, energy plus three Euclidean dimensions). This integral, like the subsequent integrals below, can be evaluated numerically using standard adaptive quadrature techniques. 

%Letting $N' = N/(k_{B}Td^{3})$, $x=\cos\theta$ and $\Xi(V') = g(V')[1-f(V',\epsilon_{F})]$, Eq. 11 becomes 
%\begin{equation}
%N' = \int \limits_{0}^{R_{1}/\bar{\alpha}} \int \limits_{-1}^{1} \int \limits_{-\infty}^{R_{1}+U'-\bar{r}(\bar{\alpha}+\beta x)} \, \Xi(V')\bar{r}^{2} \, \mathrm{d}V' \mathrm{d}x \mathrm{d}\bar{r}.
%\end{equation}
A fully disordered distribution of sites (monomer units, in our case) in Euclidean space is described by a homogeneous, isotropic site density distribution. As shown in Ref.~\cite{Apsley1975}, this corresponds to a multidimensional integral taken over a smooth, egg-shaped contour in range-space, as displayed in Fig. 2 of the main article (see also the discussion in~\cite{Apsley1975}. The asymmetry between hopping in the direction of the field (down-field) and against (up-field) is incorporated into the $\theta$-dependent upper bound of the $V'$ integral in Eq. (A11). A graphic depiction of the integral in Eq.(A11) can be found, along with further details, in~\cite{Apsley1975}.

\subsection{Estimate of bounds for $N_{inter}+N_{intra}$}
Charge-carriers can also travel along a conjugated polymer chain (intra-chain transport) before hopping to another chain. Therefore, the true number of empty sites the electron can hop to in time less than $\exp(R_{1})$ is $N_{inter+intra} > N_{inter} \equiv N$, according to the model discussed in the main article. In this paper we set this structural correction factor $N_{corr}=N_{inter+intra}/N_{inter}$ as a topological parameter which accounts for intra-chain transport and we quantitatively determine it by fitting our model to experimental data (see Fig.3 in the main article for the comparison).
%If the number of monomer units along the chain involved in the intra-chain fast transport is on the order of the number of monomer units corresponding to the persistence length, which is about 4 units, and every unit visited along the chain brings an extra hopping site which becomes accessible on nearby chains, then we estimate the following bounds on the interval: $1<N_{inter}+N_{inter}<4$.
%Interestingly, this is exactly the range where the values fitted to the experimental data fall into in Fig. 3(b) of the main article, which are all comprised between $1.5$ and $3.5$. This agreement lends further support to the accuracy of the mobility calculation. 
In this way, we can link the macroscopic mobility measurements to intra-chain transport, and its interplay/competition with inter-chain hopping. This analysis, as shown in the main article, provides insights into the role of chain morphology on the mobility.  

\subsection{Estimate of $\bar{r}_{F}$}
The next step in our derivation is to calculate $\bar{r}_{F}$, the component of the hopping vector parallel to the electric field. As there is no correlation between chain morphology and the electric field direction in the fully disordered amorphous regions, it is meaningful that $\bar{r}_{F}$ is independent of intra-chain transport. It is computed by averaging $r\cos\theta$ over all possible values of $V'$ and $\theta$, for fixed values of $\bar{R}_{nn}$, as~\cite{Apsley1975}
\begin{equation}
r\cos\theta=\bar{r}_{F}(U^{'}, T, \beta) = d\frac{I_{1}+I_{2}}{I_{3}+I_{4}}.
\end{equation}
Here $I_{1}$ and $I_{3}$ implement the first expression, while $I_{2}$ and $I_{4}$ apply the second expression for the nearest-neighbour range $\bar{R}_{nn}$ in Eq. (A9):
\begin{equation}
\begin{split}
I_{1} &= \int \limits_{-1}^{1} \int \limits_{U'-\frac{\bar{R}_{nn}\beta x}{\bar{\alpha}}}^{U'+\bar{R}_{nn}} \, \Xi(V^{'})\bigg(\frac{\bar{R}_{nn}-V'+U'}{\bar{\alpha}+\beta x}\bigg)^{3}x \, \mathrm{d}V^{'} \mathrm{d}x \\
I_{2} &= \int \limits_{-1}^{1} \int \limits_{-\infty}^{U'-\frac{\bar{R}_{nn}\beta x}{\bar{\alpha}}} \, \Xi(V^{'})\bar{R}_{nn}^{3}x \, \mathrm{d}V^{'} \mathrm{d}x \\
I_{3} &= \int \limits_{-1}^{1} \int \limits_{U'-\frac{\bar{R}_{nn}\beta x}{\bar{\alpha}}}^{U'+\bar{R}_{nn}} \, \Xi(V^{'})\bigg(\frac{\bar{R}_{nn}-V'+U'}{\bar{\alpha}+\beta x}\bigg)^{2} \, \mathrm{d}V^{'} \mathrm{d}x \\
I_{4} &= \int \limits_{-1}^{1} \int \limits_{-\infty}^{U'-\frac{\bar{R}_{nn}\beta x}{\bar{\alpha}}} \, \Xi(V^{'})\bar{R}_{nn} \, \mathrm{d}V^{'} \mathrm{d}x, \\
\end{split}
\end{equation}
where $x=\cos\theta$ and $\Xi(V') = g(V')[1-f(V',\epsilon_{F})]$.

\subsection{Disorder-averaged mobility}
The final step in our calculation of the microscopic inter-chain hopping mobility of charge-carriers in strongly disordered regions of conjugated polymers, as given by Eq. (A1), is to average over all site energies $U'$ using the electronic DOS given by Eq. (A2):
\begin{equation}
\begin{split}
\mu_{dis}(T,\beta)&=\frac{k_{B}T \int \limits_{-\infty}^{\infty} \, g(U')f(U',\epsilon_{F})\mu_{dis}(U',T,\beta) \, \mathrm{d}U'}{k_{B}T \int \limits_{-\infty}^{\infty} \, g(U')f(U',\epsilon_{F}) \, \mathrm{d}U'}\\
\
&=\frac{k_{B}T}{n_{c}} \int \limits_{-\infty}^{\infty} \, g(U')f(U',\epsilon_{F})\mu_{dis}(U',T,\beta) \, \mathrm{d}U'.
\end{split}
\end{equation}

Working backwards this parameter can be used to retrieve the macroscopic observable mobility of a semi-crystalline, conjugated polymer sample, as a function of the characteristic experimental system variables; the temperature $T$ and the dimensionless electric field parameter $\beta$,
\begin{equation}
\mu(T,\beta)\simeq \frac{L_{p}\mu_{dis}(T,\beta)}{d_{a}}=\frac{\mu_{dis} (T,\beta)}{1-x}.
\end{equation}
where $x=0.65$ in accordance with the experimental characterization of the sample.
\end{appendix}

%\begin{acknowledgment}
\section*{Acknowledgements}
The authors would like to thank Prof. Dr. Heinz B\"{a}ssler for helpful discussions and suggestions. This study was supported by the Winton Programme for the Physics of Sustainability (B.O.C.).
%\end{acknowledgment}

\end{document}